\documentclass[useAMS,usenatbib]{mn2e}
\usepackage{graphicx}
\begin{document}

\title{Testing the rotating hot spot model using X-ray burst oscillations from 4U~1636$-$536}

\author[R. Artigue et al.]{Romain Artigue$^{1,2}$\thanks{E-mail: romain.artigue@irap.omp.eu}, Didier Barret$^{1,2}$, Frederick K. Lamb$^{3,4}$, Ka Ho Lo$^3$, \and M. Coleman Miller$^5$\\
$^1$Universit\'e de Toulouse; UPS-OMP; IRAP; Toulouse, France.\\
$^2$CNRS; Institut de Recherche en Astrophysique et Plan\'etologie, 9 Avenue du Colonel Roche, 31028, Toulouse, France.\\
$^3$Center for Theoretical Astrophysics and Department of Physics, University of Illinois at Urbana-Champaign, 1110 West Green Street, \\Urbana, IL 61801-3080, USA.\\
$^4$Department of Astronomy, University of Illinois at Urbana-Champaign, 1002 West Green Street, Urbana, IL 61801-3074, USA.\\
$^5$University of Maryland, Department of Astronomy and Joint Space-Science Institute, College Park, MD 20742-2421, USA.}

\maketitle

\begin{abstract}
Precise and accurate measurements of neutron star masses and radii would provide valuable information about the still uncertain properties of cold matter at supranuclear densities. One promising approach to making such measurements involves analysis of the X-ray flux oscillations often seen during thermonuclear (type~1) X-ray bursts. These oscillations are almost certainly produced by emission from hotter regions on the stellar surface modulated by the rotation of the star. One consequence of the rotation is that the oscillation should appear earlier at higher photon energies than at lower energies. \citet{1999ApJ...519L..73F} found compelling evidence for such a hard lead in the tail oscillations of one type~1 burst from Aql~X-1. Subsequently, \citet*{2003ApJ...595.1066M} analyzed oscillations in the tails of type~1 bursts observed using \textit{RXTE}. They found significant evidence for variation of the oscillation phase with energy in 13 of the 51 oscillation trains they analyzed and an apparent linear trend of the phase with energy in six of nine average oscillation profiles produced by folding the energy-resolved oscillation waveforms from five stars and then averaging them in groups. In four of these nine averaged energy-resolved profiles, the oscillation appeared to arrive earlier at lower energies than at higher energies. Such a trend is inconsistent with a simple rotating hot spot model of the burst oscillations and, if confirmed, would mean that this model cannot be used to constrain the masses and radii of these stars and would raise questions about its applicability to other stars. We have therefore re-analyzed individually the oscillations observed in the tails of the four type~1 bursts from 4U~1636$-$536 that, when averaged, provided the strongest evidence for a soft lead in the analysis by \citet{2003ApJ...595.1066M}. We have also analyzed the oscillation observed during the superburst from this star. We find that the data from these five bursts, treated both individually and jointly, are fully consistent with a rotating hot spot model. Unfortunately, the uncertainties in these data are too large to provide interesting constraints on the mass and radius of this star.
\end{abstract}

\section{Introduction}
The central regions of neutron stars contain highly degenerate matter with densities up to several times nuclear saturation density. The properties of this matter cannot be studied in terrestrial laboratories but can be probed by determining the masses and radii of neutron stars \citep[see, e.g.,][]{2007PhR...442..109L}. One method proposed to obtain this information involves fitting detailed models to the waveforms of the oscillations often seen during type~1 (thermonuclear) X-ray bursts (see, e.g., \citealt{1992ApJ...388..138S,1998ApJ...499L..37M,2000ApJ...531..447B,2001ApJ...546.1098W,2002ApJ...564..353N,2004AIPC..714..245S,2005ApJ...618..451C,2005ApJ...619..483B,2007ApJ...654..458C,2013ApJ...762...96B}; see \citealt{2012ARA&A..50..609W} for a recent overview of the properties of these bursts). These oscillations are thought to be produced by rotational modulation of the emission from a hot spot that rotates at or close to the stellar spin frequency \citep{1996ApJ...469L...9S}. The amplitude of the oscillation and its harmonic content are affected by special relativistic Doppler boosts and aberration and gravitational light deflection, which depend on the mass and radius of the star. The mass and radius of the star can therefore be determined by fitting sufficiently accurate waveform models to waveform data of sufficient quality.

A strong prediction of the rotating hot spot model is that the oscillation should appear earlier at higher photon energies than at lower energies. This is most easily seen in an extreme case. The spectral shape of the emission from burst atmospheres typically has a shape close to the shape of a Bose-Einstein spectrum \citep[see, e.g.,][]{2010ApJ...720L..15B}. At energies well above the peak of the spectrum, the observed maximum of the burst oscillation will occur at the stellar rotational phase that maximizes the blueshift of the hot spot relative to the observer, which happens when the spot is near the approaching limb of the star. In contrast, at energies near the peak of the spectrum the maximum of the oscillation will occur later, when the projected area of the spot is close to its maximum. \citet{1999ApJ...519L..73F} found just such a hard lead in \textit{Rossi X-ray Timing Explorer} (\textit{RXTE}) observations of the oscillation during the tail of a type~1 burst from Aql~X-1.

\citet*{2003ApJ...595.1066M} subsequently analyzed 51 oscillation trains in \textit{RXTE} data from the tails of type~1 bursts from six neutron stars. They folded each of these oscillation trains to produce oscillation profiles and then measured the phases of these profiles using linear least-squares fits of sinusoids to the oscillation profiles. Although the uncertainties of the oscillation phases were relatively large, they found significant evidence for phase variations with photon energy in 13 of the 51 oscillation trains. Seven of these 13 trains showed random phase variations while the remaining six showed either no significant lags or a roughly linear phase trend corresponding to a hard lag. 

\citet{2003ApJ...595.1066M} then grouped all the energy-resolved oscillation profiles from a given star for each \textit{RXTE} PCA gain epoch and averaged the profiles within each of the resulting nine groups. They found that five of the nine averaged  energy-resolved oscillation profiles showed significant phase variations with energy. The phase variations in one of these five profiles appeared to be random while the phase variations in the other four were consistent with a linear increase in phase with energy. The apparent hard lags in these four profiles are inconsistent with a simple rotating hot spot model of burst oscillations. \citeauthor{2003ApJ...595.1066M} speculated that these hard lags might be caused by Comptonization. If confirmed, statistically significant hard lags would mean that a simple rotating spot model of burst oscillations cannot be used to constrain the masses and radii of these stars and would raise questions about the applicability of this model to other stars.

Here we re-analyze individually the oscillations observed in the tails of four type~1 bursts from 4U~1636$-$536. The average oscillation profiles \citet{2003ApJ...595.1066M} constructed by summing the individual profiles from these bursts provided the strongest evidence for a soft lead in their burst oscillation analysis. We also analyze the oscillation observed during the superburst from this star. In contrast to \citeauthor{2003ApJ...595.1066M}, we find that the data from these bursts, treated both individually and jointly, are fully consistent with a rotating hot spot model. The uncertainties in the \textit{RXTE} data are too large to provide interesting constraints on the mass and radius of this star.

The number of counts in these burst oscillation profiles are so small that the consistency of these profiles with the rotating spot model cannot be taken as confirmation of this model. However, this consistency does mean that these data do not challenge the rotating spot model and that fitting this model to burst oscillation data remains a viable way to constrain neutron star masses and radii using higher quality data from future large-area timing missions such as 
the Neutron Star Interior Composition ExploreR 
\citep[\textit{NICER};][]{2012SPIE.8443E..13G}, 
the Large Observatory for X-ray Timing 
\citep[\textit{LOFT};][]{mign12, delm12}, and 
the Advanced X-ray Timing Array \citep[\textit{AXTAR};][]{chak08, ray11}.

In Section~\ref{sec:data-analysis-results}, we describe the X-ray data we use in this study and the results of our timing analysis. In Section~\ref{sec:discussion}, we discuss our results and conclusions.

\section{Data, analysis, and results}
\label{sec:data-analysis-results}

Our analysis is based on Event Mode data from the \textit{RXTE} Proportional Counter Array (PCA). This mode has a time resolution of 1/8192 seconds and is therefore well-suited to studying burst oscillations with frequencies of hundreds of Hertz. We focused on the 4U~1636$-$536 type~1 bursts observed during PCA gain Epoch~4 (the boundaries of the PCA energy channels are slightly different in different epochs), because the averaged energy-resolved folded profile of the oscillations during these bursts that was constructed by \citet{2003ApJ...595.1066M} appeared to show a systematic soft lead, contrary to what is predicted by a simple rotating hot spot model (see the top middle panel of their Figure~2). We also analyzed data from \textit{RXTE} observations of the hours-long superburst from 4U~1636$-$536 \citep{2002ApJ...577..337S}, which occurred during PCA gain Epoch~5, to determine whether we could obtain an acceptable joint fit to the data on all these bursts.

There were four type~1 bursts rom 4U~1636$-$536 during PCA gain Epoch~4 that had detectable oscillations; their ObsIDs are listed in the first four rows of Table~1. Epoch~4 ObsIDs 40028-01-08-00 and 40031-01-01-06 each had one additional oscillation train some seconds after the segments we analyzed, but these trains were too weak to satisfy our significance criterion and hence we have not included them in our analysis.

We followed \citet{2003ApJ...595.1066M} by using five energy bins spanning photon energies from 2 and 23 keV, which corresponds roughly to PCA energy channels 5 through 53 (of the 256 total channels). We also followed \citeauthor{2003ApJ...595.1066M} by analyzing the oscillations during the tails of the bursts rather than during their rising portions, where the changing frequencies and rapidly changing amplitudes of the oscillation trains could introduce additional complications.

\begin{table*}
\centering
\begin{minipage}{140mm}
\caption{Properties of the 4U~1636$-$536 burst segments analyzed}
\begin{tabular}{@{}rrrr@{}}
\hline
ObsID & Start time & Duration (s) & Mean Frequency (Hz)\\
\hline
40028-01-06-00 & 171611735.014 & 3.260 & 580.3985\\
40028-01-08-00 & 172366987.063 & 1.301 & 580.5576\\
40030-03-04-00 & 172431061.302 & 1.926 & 580.4122\\
40031-01-01-06 & 172609556.020 & 2.645 & 581.1129\\
50030-02-08-01 & 225479593.957 & 149.251 & 581.9692\\
\hline
\end{tabular}
\label{table:burst-segments}
\end{minipage}
\end{table*}

For each burst, we first searched for the starting time and duration of the segment of the burst that maximized the significance of the oscillating signal. Searches using the Leahy power \citep{1983ApJ...266..160L} or, equivalently, the $Z_1^2$ statistic \citep{1983A&A...128..245B, 1999ApJ...516L..81S}, and the deviation of $\chi^2$ for the best folded profile from its value assuming no oscillation all yielded equivalent results. For the four type~1 bursts, we explored durations ranging from 0.5 to 6~seconds in steps of 1/16 of a second, with starting times 1/16 of a second apart throughout the burst tail. For the superburst, we tried durations between 50 and 250 seconds in steps of 4 seconds throughout the portion of the burst where X-ray flux oscillations were observed. For the first type~1 burst and the superburst, we also tried fits with a frequency that varies linearly in time; the results were indistinguishable from those for our fits assuming a constant frequency. We then used epoch folding \citep[see, e.g.,][]{1990MNRAS.244...93D} to determine the frequency of the oscillation during each burst segment and to construct a folded oscillation profile for each set of energy channels. The start time, duration, and oscillation frequency for each of the five burst segments we analyzed are listed in Table~\ref{table:burst-segments}.

We determined the zero phase (maximum flux) of the bolometric oscillation profile and the oscillation profile in each set of energy channels by fitting a sinusoidal model to each of these profiles. The parameters in this model were the phase, amplitude, and DC level. We estimated the $1\sigma$ uncertainties in these parameters using the standard approach. Namely, we began with the overall best-fit values of the three parameters. We then varied one parameter from its best-fit value while minimizing $\chi^2$ with respect to the other two parameters, until we found the value of the first parameter that gave a $\Delta\chi^2$ of 1 relative to the $\chi^2$ for the best-fit parameter values. We used this value of the first parameter to define its $1\sigma$ uncertainty region. We then repeated this procedure for the other two parameters.

Following \citet{2003ApJ...595.1066M}, we used the parameter values in our fitted sinusoidal models to determine the phase of the oscillation profile in each set of energy channels relative to the phase of the bolometric oscillation profile. We also computed the relative phases by cross-correlating the profiles. The results were consistent with the results from the sinusoidal fits. Both methods confirmed the hard lead found by \citet{1999ApJ...519L..73F} for the oscillation in a segment of the tail of a burst from Aql~X-1. We also studied earlier portions of that burst but found large data gaps that render any conclusions concerning these portions untrustworthy.

\begin{figure*}
\includegraphics[scale=0.7]{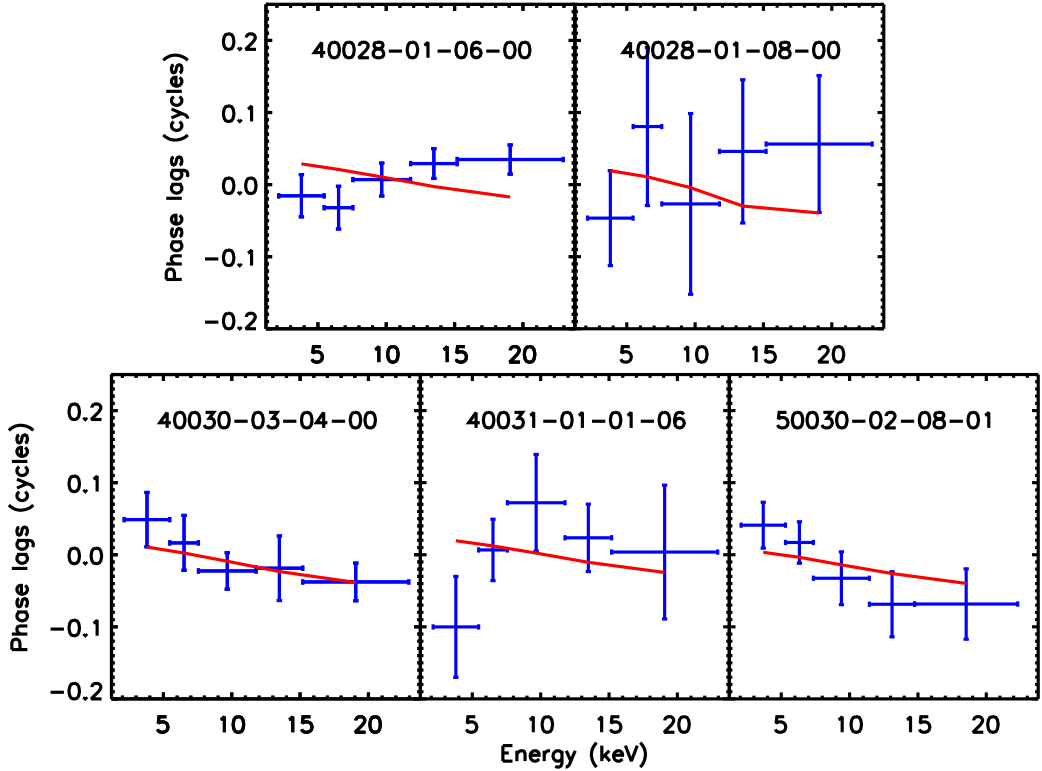}
\caption{Phase lags for oscillations in the tails of five X-ray bursts from 4U~1636$-$536 (data points) and a joint fit of a simple rotating circular hot spot model to all the data (broken lines), showing that this model is fully consistent with the data. The phases shown for the data points and the model points are both relative to the zero phase of the observed bolometric oscillation profile, which is defined by the maximum flux of the profile. Each panel shows the data and ObsID for one of the five bursts. The horizontal error bars indicate the approximate energy range covered by each channel while the vertical error bars show the $1\sigma$ uncertainties determined by fitting a sinusoid to the data. In fitting the rotating spot model to these data, the mass, radius, observer latitude, and distance to the star were kept fixed for all five bursts, but the spot latitude, spot radius, and spot color temperature were allowed to vary from burst to burst. See the text for further details.
}
\label{fig:lags}
\end{figure*}

Figure~\ref{fig:lags} shows our results for the phases of the oscillation profiles as a function of photon energy. These results are insensitive to small changes in the start times and durations of the data segment and the oscillation frequencies. In this figure, a negative lag implies that the maximum of the oscillation in that energy range arrived, on average, before the maximum of the oscillation in the bolometric profile. A positive lag means the opposite. As explained earlier, a simple rotating spot model predicts an increasingly negative lag with increasing energy.

Figure~\ref{fig:lags} also shows the results of a fit to these data of the model waveform produced by a uniform circular hot spot fixed to the rotating star, computed using the algorithms described briefly in \citet{2009ApJ...706..417L,2009ApJ...705L..36L} and more extensively in Lo et al.\ (2013, in preparation). In fitting this model, the mass, radius, observer latitude, and distance to the star were kept fixed for all five bursts, but the latitude of the spot center, the spot angular radius, and the spot color temperature were allowed to be different for each burst. In order to extract maximal information from the data, we used standard Bayesian techniques to compare our model predictions with the data, which were broken into 16 phases and 25 energy channel groupings per phase for each burst. The fit shown is for a star with a mass of $1.5~M_\odot$ and a radius of 8.8~km, an observer inclination relative to the stellar spin axis of $56^\circ$, and a distance of 6.5~kpc (consistent with the distance of $(6.0\pm 0.5)(M/1.4~M_\odot)^{1/2}~{\rm kpc}$ estimated by \citealt{2006ApJ...639.1033G}). A wide range of other stellar masses and radii give comparably good fits to these data, so we are unable to derive interesting constraints on the properties of this neutron star using these data.

A simple rotating hot spot model is fully consistent with the 4U~1636$-$536 data that we analyzed. The fits of this model go through 76\% (19 of 25) of the one-sigma error regions. The Cash statistic (\citealt{1979ApJ...228..939C}; note that this statistic is an analog of $\chi^2$ for small numbers of counts and asymptotes to $\chi^2$ for large numbers of counts) for the joint fit is 1859 for 1850 degrees of freedom. If we treat this as a $\chi^2$ distribution, then if the model is correct we will find $\chi^2$ of 1859 or larger for 1850 degrees of freedom in approximately 44\% of realizations. This is therefore an excellent fit, especially because the data shown for each burst were obtained by folding the oscillation waveform over several seconds of a burst that itself lasted only a few seconds, a procedure that undoubtedly smeared the oscillation profile. The uncertainties in the lag estimates are sufficiently large that occasional apparent soft leads (such as the one in the leftmost panel) are consistent with being statistical fluctuations. The superburst, which has far more counts than any of the type~1 bursts, does appear to show a hard lead, in accordance with the prediction of a simple rotating spot model.  

\newpage
\section{Discussion}\label{sec:discussion}

We have individually analyzed the X-ray flux oscillations during four type~1 X-ray bursts from 4U~1636$-$536 observed using \textit{RXTE} during its PCA gain Epoch 4 and the superburst from this star. We find that the variations of the phases of these oscillations with photon energy are fully consistent with a rotating hot spot model, whether the data from each burst is fit individually or jointly. Unfortunately, the uncertainties in these data are too large to provide interesting constraints on the mass and radius of this star.

Our results for the variations of the phases of the oscillation profiles with photon energy differ from the results reported by \citet{2003ApJ...595.1066M}, which appeared to be inconsistent with a simple rotating hot spot model. The analysis by \citeauthor{2003ApJ...595.1066M} differed from ours in several respects. In particular, \citeauthor{2003ApJ...595.1066M} averaged the energy-resolved profiles of the oscillations observed during the type~1 X-ray bursts from 4U~1636$-$536 during \textit{RXTE} PCA gain Epoch 4 and then analyzed them, whereas we analyzed the energy-resolved profiles of the oscillations observed during each burst individually and then jointly. 

Averaging the oscillation profiles before analyzing them is not the optimal procedure, because the variations of the phases of the oscillation profiles with energy differ from burst to burst in this data set, as Figure~\ref{fig:lags} shows. This is to be expected, because the properties of the hot spot (such as its radius or inclination from the spin axis) can change from burst to burst. Indeed, \citeauthor{2003ApJ...595.1066M} found that the variations with energy of the phases of the oscillation profiles in the  4U~1636$-$536 bursts changed with time (the dependence during PCA gain Epochs 3, 4, and 5 differed from one another). This suggests that the oscillation profile does vary from burst to burst.

Our most important result is that the X-ray flux oscillations during the four type~1 X-ray bursts from 4U~1636$-$536 that we analyzed and the superburst from this star are fully consistent with a simple rotating hot spot model, whether the data from each burst are fit individually or jointly. Thus, these data do not call into question the use of such a model to fit X-ray burst oscillation data and thereby constrain the masses and radii of individual neutron stars. At the same time, it is clear that more and better data will be required to obtain tight constraints.

Proposed space missions that will obtain much better timing data include \textit{NICER}, which will focus on deep observations of X-ray emitting millisecond pulsars (\citealt{2012SPIE.8443E..13G}; see \citealt{2013ApJ...762...96B} for constraints recently obtained using \textit{XMM-Newton} data); \textit{LOFT}, which will have a collecting area more than an order of magnitude larger than the \textit{RXTE} PCA \cite[see][]{mign12, delm12}; and \textit{AXTAR}, which would also have a collecting area much larger than the \textit{RXTE} PCA \citep{chak08, ray11}. Our simulations of the precision of the phase lag measurements that could be achieved using \textit{LOFT} show that it could achieve a precision of $\sim\!0.01$~cycles, compared to the $\sim\!0.05$ precision achieved using \textit{RXTE}. A full Bayesian analysis of the constraints on the masses and radii of neutron stars that could be achieved by fitting waveform models to \textit{LOFT} observations of burst oscillations (Lo et al. 2013, in preparation) indicates that tight constraints can be achieved for systems that have a favorable geometry. Thus, there is a good prospect that these missions will provide neutron star mass and radius estimates precise enough to tightly constrain the properties of cold supranuclear matter.

\section{Acknowledgments}
We thank Mike Muno for carefully reading and commenting on this manuscript.
This work was supported in part by a grant from the Simons Foundation (grant number 230349 to MCM), NSF grant AST0708424 at Maryland, and NSF grant AST0709015 and the Fortner Chair at Illinois.  MCM also thanks the Department of Physics and Astronomy at Johns Hopkins University for their hospitality during his sabbatical.

\bibliography{ms}

\begin{thebibliography}{}

\bibitem[\protect\citeauthoryear{{Bhattacharyya}, {Strohmayer}, {Miller} \&
  {Markwardt}}{{Bhattacharyya} et~al.}{2005}]{2005ApJ...619..483B}
{Bhattacharyya} S.,  {Strohmayer} T.~E.,  {Miller} M.~C.,    {Markwardt} C.~B.,
   2005, \apj, 619, 483

\bibitem[\protect\citeauthoryear{{Bogdanov}}{{Bogdanov}}{2013}]{2013ApJ...762...96B}
{Bogdanov} S.,  2013, \apj, 762, 96

\bibitem[\protect\citeauthoryear{{Boutloukos}, {Miller} \& {Lamb}}{{Boutloukos}
  et~al.}{2010}]{2010ApJ...720L..15B}
{Boutloukos} S.,  {Miller} M.~C.,    {Lamb} F.~K.,  2010, \apjl, 720, L15

\bibitem[\protect\citeauthoryear{{Braje}, {Romani} \& {Rauch}}{{Braje}
  et~al.}{2000}]{2000ApJ...531..447B}
{Braje} T.~M.,  {Romani} R.~W.,    {Rauch} K.~P.,  2000, \apj, 531, 447

\bibitem[\protect\citeauthoryear{{Buccheri} et~al.,}{{Buccheri}
  et~al.}{1983}]{1983A&A...128..245B}
{Buccheri} R.  et~al., 1983, \aap, 128, 245

\bibitem[\protect\citeauthoryear{{Cadeau}, {Leahy} \& {Morsink}}{{Cadeau}
  et~al.}{2005}]{2005ApJ...618..451C}
{Cadeau} C.,  {Leahy} D.~A.,    {Morsink} S.~M.,  2005, \apj, 618, 451

\bibitem[\protect\citeauthoryear{{Cadeau}, {Morsink}, {Leahy} \&
  {Campbell}}{{Cadeau} et~al.}{2007}]{2007ApJ...654..458C}
{Cadeau} C.,  {Morsink} S.~M.,  {Leahy} D.,    {Campbell} S.~S.,  2007, \apj,
  654, 458

\bibitem[\protect\citeauthoryear{{Cash}}{{Cash}}{1979}]{1979ApJ...228..939C}
{Cash} W.,  1979, \apj, 228, 939

\bibitem[\protect\citeauthoryear{Chakrabarty, Ray \& Strohmayer}{Chakrabarty
  et~al.}{2008}]{chak08}
Chakrabarty D.,  Ray P.~S.,    Strohmayer T.~E.,  2008, in {A Decade of
  Accreting Millisecond X-ray Pulsars}. pp 227--230

\bibitem[\protect\citeauthoryear{{Davies}}{{Davies}}{1990}]{1990MNRAS.244...93D}
{Davies} S.~R.,  1990, \mnras, 244, 93

\bibitem[\protect\citeauthoryear{Del~Monte, Donnarumma \& Consortium}{Del~Monte
  et~al.}{2012}]{delm12}
Del~Monte E.,  Donnarumma I.,    Consortium L.,  2012, Memorie della Societa
  Astronomica Italiana, 83, 352

\bibitem[\protect\citeauthoryear{{Ford}}{{Ford}}{1999}]{1999ApJ...519L..73F}
{Ford} E.~C.,  1999, \apjl, 519, L73

\bibitem[\protect\citeauthoryear{{Galloway}, {Psaltis}, {Muno} \&
  {Chakrabarty}}{{Galloway} et~al.}{2006}]{2006ApJ...639.1033G}
{Galloway} D.~K.,  {Psaltis} D.,  {Muno} M.~P.,    {Chakrabarty} D.,  2006,
  \apj, 639, 1033

\bibitem[\protect\citeauthoryear{{Gendreau}, {Arzoumanian} \&
  {Okajima}}{{Gendreau} et~al.}{2012}]{2012SPIE.8443E..13G}
{Gendreau} K.~C.,  {Arzoumanian} Z.,    {Okajima} T.,  2012, in Society of
  Photo-Optical Instrumentation Engineers (SPIE) Conference Series.

\bibitem[\protect\citeauthoryear{{Lamb}, {Boutloukos}, {Van Wassenhove},
  {Chamberlain}, {Lo}, {Clare}, {Yu} \& {Miller}}{{Lamb}
  et~al.}{2009a}]{2009ApJ...706..417L}
{Lamb} F.~K.,  {Boutloukos} S.,  {Van Wassenhove} S.,  {Chamberlain} R.~T.,
  {Lo} K.~H.,  {Clare} A.,  {Yu} W.,    {Miller} M.~C.,  2009, \apj, 706, 417

\bibitem[\protect\citeauthoryear{{Lamb}, {Boutloukos}, {Van Wassenhove},
  {Chamberlain}, {Lo} \& {Miller}}{{Lamb} et~al.}{2009b}]{2009ApJ...705L..36L}
{Lamb} F.~K.,  {Boutloukos} S.,  {Van Wassenhove} S.,  {Chamberlain} R.~T.,
  {Lo} K.~H.,    {Miller} M.~C.,  2009, \apjl, 705, L36

\bibitem[\protect\citeauthoryear{{Lattimer} \& {Prakash}}{{Lattimer} \&
  {Prakash}}{2007}]{2007PhR...442..109L}
{Lattimer} J.~M.,  {Prakash} M.,  2007, \physrep, 442, 109

\bibitem[\protect\citeauthoryear{{Leahy}, {Darbro}, {Elsner}, {Weisskopf},
  {Kahn}, {Sutherland} \& {Grindlay}}{{Leahy}
  et~al.}{1983}]{1983ApJ...266..160L}
{Leahy} D.~A.,  {Darbro} W.,  {Elsner} R.~F.,  {Weisskopf} M.~C.,  {Kahn} S.,
  {Sutherland} P.~G.,    {Grindlay} J.~E.,  1983, \apj, 266, 160

\bibitem[\protect\citeauthoryear{Mignani et~al.,}{Mignani
  et~al.}{2012}]{mign12}
Mignani R.~P.  et~al., 2012, in New Horizons in Time-Domain Astronomy. pp
  372--375

\bibitem[\protect\citeauthoryear{{Miller} \& {Lamb}}{{Miller} \&
  {Lamb}}{1998}]{1998ApJ...499L..37M}
{Miller} M.~C.,  {Lamb} F.~K.,  1998, \apjl, 499, L37

\bibitem[\protect\citeauthoryear{{Muno}, {{\"O}zel} \& {Chakrabarty}}{{Muno}
  et~al.}{2003}]{2003ApJ...595.1066M}
{Muno} M.~P.,  {{\"O}zel} F.,    {Chakrabarty} D.,  2003, \apj, 595, 1066

\bibitem[\protect\citeauthoryear{{Nath}, {Strohmayer} \& {Swank}}{{Nath}
  et~al.}{2002}]{2002ApJ...564..353N}
{Nath} N.~R.,  {Strohmayer} T.~E.,    {Swank} J.~H.,  2002, \apj, 564, 353

\bibitem[\protect\citeauthoryear{Ray, Phlips, Wood, Chakrabarty, Remillard \&
  Wilson-Hodge}{Ray et~al.}{2011}]{ray11}
Ray P.~S.,  Phlips B.~F.,  Wood K.~S.,  Chakrabarty D.,  Remillard R.~A.,
  Wilson-Hodge C.~A.,  2011, in Fast X-ray timing and spectroscopy at extreme
  count rates (HTRS 2011).

\bibitem[\protect\citeauthoryear{{Strohmayer}}{{Strohmayer}}{1992}]{1992ApJ...388..138S}
{Strohmayer} T.~E.,  1992, \apj, 388, 138

\bibitem[\protect\citeauthoryear{{Strohmayer}}{{Strohmayer}}{2004}]{2004AIPC..714..245S}
{Strohmayer} T.~E.,  2004, in {Kaaret} P.,  {Lamb} F.~K.,   {Swank} J.~H.,
  eds,  American Institute of Physics Conference Series Vol. 714, X-ray Timing
  2003: Rossi and Beyond. pp 245--252

\bibitem[\protect\citeauthoryear{{Strohmayer} \& {Markwardt}}{{Strohmayer} \&
  {Markwardt}}{1999}]{1999ApJ...516L..81S}
{Strohmayer} T.~E.,  {Markwardt} C.~B.,  1999, \apjl, 516, L81

\bibitem[\protect\citeauthoryear{{Strohmayer} \& {Markwardt}}{{Strohmayer} \&
  {Markwardt}}{2002}]{2002ApJ...577..337S}
{Strohmayer} T.~E.,  {Markwardt} C.~B.,  2002, \apj, 577, 337

\bibitem[\protect\citeauthoryear{{Strohmayer}, {Zhang}, {Swank}, {Smale},
  {Titarchuk}, {Day} \& {Lee}}{{Strohmayer} et~al.}{1996}]{1996ApJ...469L...9S}
{Strohmayer} T.~E.,  {Zhang} W.,  {Swank} J.~H.,  {Smale} A.,  {Titarchuk} L.,
  {Day} C.,    {Lee} U.,  1996, \apjl, 469, L9

\bibitem[\protect\citeauthoryear{{Watts}}{{Watts}}{2012}]{2012ARA&A..50..609W}
{Watts} A.~L.,  2012, \araa, 50, 609

\bibitem[\protect\citeauthoryear{{Weinberg}, {Miller} \& {Lamb}}{{Weinberg}
  et~al.}{2001}]{2001ApJ...546.1098W}
{Weinberg} N.,  {Miller} M.~C.,    {Lamb} D.~Q.,  2001, \apj, 546, 1098

\end{thebibliography}
\end{document}